\documentclass[twocolumn,showpacs,aps]{revtex4}   
\usepackage{epsf}

\def\hb{\hbar}

\def\tO{{{\tilde \Omega}}}   
\def\ep{\varepsilon}   
\def\lep{{|\mathrm{log }\ \ep|}}

\def\O{\Omega}   
   
\def\a0{\alpha_0}   
\def\a{\alpha}   
\def\b{\beta}   
\def\g{\gamma}   
\def\tg{{\tilde{\gamma}}}   
   
\def\rtf{\rho}   
   
\def\r0{\rho_{0}}

\def\be{\begin{equation}}   
\def\ee{\end{equation}}   
\def\beq{\begin{equation}}   
\def\eeq{\end{equation}}

\def\cd{{\cal D}}

\def\ty{{\tilde y}}   
\def\tz{{\tilde z}}

\newcommand{\zmax}{ z_{\mbox{\scriptsize{max}}} }

\newcommand{\calD}{{\mathcal{D}}}

\newtheorem{theo}{Theorem}   
\newtheorem{prop}{Proposition}   
\newtheorem{remark}{Remark}   
\begin{document}   
   
\title{On the shape of vortices for a rotating Bose Einstein condensate}   
   
\author{Amandine Aftalion}   
\email{aftalion@ann.jussieu.fr}   
\affiliation{CNRS and Laboratoire Jacques-Louis Lions,  Universit\'e Paris 6, 175 rue du Chevaleret, 75013 Paris,   
France.}   
\author{Robert L. Jerrard}   
\email{rjerrard@math.uiuc.edu}   
\affiliation{Mathematics Department, University of Illinois at Urbana Champaign, Urbana, IL 61801, USA.}   
\date{\today}

             
\begin{abstract}   
For a Bose-Einstein condensate placed in a rotating trap,   
 we study the simplified energy of a vortex line derived in \cite{AR}  
in order to determine the shape of the vortex line according to  
the rotational velocity and the elongation of the condensate.  The energy  reflects the competition between the  length of the vortex which needs to be minimized taking into account the anisotropy of the trap and the rotation term which pushes the vortex along the $z$ axis. We prove that if the condensate has the shape of a pancake, the vortex stays straight along the $z$ axis while in the case of a cigar, the vortex is bent. 
\end{abstract}   
   
\maketitle

\section{Introduction}   
   
Dilute  Bose-Einstein condensates have recently been achieved in   
confined alkali-metal gases  and the study of vortices is one  
of the key issues. One type of experiments  consists in imposing  
a laser beam on the magnetic trap holding the atoms to create a  
harmonic anisotropic rotating potential \cite{MCWD,MCWD2,AK,RK}.  
 Vortices are nucleated and the number of vortices depends on the rotational velocity. 
 It has been observed experimentally \cite{MCWD}, that when the first vortex is nucleated,   
the contrast is not $100\%$    
which means that the vortex line is not straight but bending.  
 Numerical computations of the Gross Pitaevskii  
equation have shown evidence of vortex bending \cite{GP1}.   
   
The aim of this paper is to   
characterize the dependence of the   
shape of the vortex  
line on the elongation of the trap and the rotational  
velocity. In particular, using a simplified energy for a vortex 
line derived in \cite{AR} from the Gross Pitaevskii energy, we study the stability and instability of the straight vortex and 
we 
prove that when the condensate has a cigar shape the first 
vortex is bent, while when it is a pancake, the first vortex 
is straight and lies on the axis of rotation.  
We also show 
that  
vortices cannot be nucleated too close to the boundary, because they 
have a minimal length.

In \cite{AR}, we have derived  a simplified expression for the 
energy of several vortex lines in a rotating trap from the usual 
Gross Pitaevskii energy   describing the steady state of the condensate 
\begin{eqnarray}  
\label{BE}  
{\cal E}_{3D}(\phi) &=& \int {\hb^2 \over{2m}} |\nabla \phi|^2+{{\hb {\tO}}}\cdot  
(i\phi, \nabla \phi\times {\bf x}) \nonumber\\  
&&+{m\over 2}  
 \sum_{\alpha} \omega^2_\alpha  r^2_\alpha |\phi|^2 +{N\over 2} g_{3D}|\phi|^4 .  
\end{eqnarray}  
We let $d=(\hb/m\omega_y)^{1/2}$ be the characteristic length,   
$\omega_x=\a\omega_y$, $\omega_z=\b\omega_y$. We define a small 
non-dimensional parameter $\ep$ which characterizes the fact that 
we are in the Thomas Fermi regime by  
$$\ep^2\sqrt{\ep}={{d}\over {4\pi Na}},$$    
where $N$ is the number of particles and $a$ the scattering 
length. In the ENS experiment \cite{MCWD2,MCWD}, $\ep=1.74\ 10^{-2}$ 
while in the MIT experiment \cite{RK}, $\ep=3.52\ 10^{-3}$. We 
rescale distances by $d/\sqrt\ep$ and the chemical potential $\mu_0$ 
so that the new chemical potential $\r0$ is given  by  
\beq\label{chem}  
\r0=2\ep{{\mu_0}\over {\hb \omega_y}}.  
\eeq  
In these units, we have $\r0=0.42$ and $\r0=0.46$ respectively 
for the ENS and MIT experiments.   
 We let   
\beq\label{rho} 
\rtf({\bf r})=\r0 -(\a^2x^2+y^2+\b^2z^2)\eeq 
 be the Thomas Fermi limit of the wave function in rescaled units. Then, we have obtained in  
  \cite{AR} a simplified expression for the energy of a vortex line $\g$, which is  
$$\ep \hb \omega_y\pi \lep \ E[\g]$$  
with  
\beq\label{Eg}   
E[\g]=\int_\g \rtf \ dl-\O \int_\g \rtf^2 \ dz,   
\eeq   
where $\O$ is related to the experimental rotational velocity $\tO$ by  
\beq\label{O}   
\O={{\tO}\over {\omega_y}} {1\over {(1+\a^2)\ep\lep}}.  
\eeq   
The energy $E[\g]$ reflects the competition between the vortex energy due to its length (1st term) and the rotation term. Note that the rotation term is an oriented integral ($dz$ not $dl$), which actually forces the vortex to be along the $z$ axis, while the other term wants to minimize the length. This is why, according to the geometry of the trap, the shape of the vortex varies. 
 
This energy is similar to the one obtained in \cite{SF2} in the study of the dynamics of the vortex line. 
Note that the energy that we actually derive in \cite{AR} is 
slightly more involved than (\ref{Eg}). In the regime of the experiments, it 
is reasonable to restrict to this expression (\ref{Eg}), taking into 
account the fact that the vortex core is sufficiently small (it 
is of size $\ep$ in our units) and neglecting  
the interaction of the curve with itself. We are interested only 
in the presence of the first vortex: when there are several 
vortices, the energy has an extra term due to the repulsion 
between the lines.   
  
In this scaling, the energy of the vortex free solution is zero. 
Thus, a vortex line is energetically favorable when $\O, \b$ are such that $\inf_\g {E}[\g]<0$. The aim of this paper is to study the shape of the vortex lines $\g$ minimizing ${E}[\g]$. 
We define the domain ${\cal D}=\{ \rho>0\}$. This is the domain 
where the condensate lies. All the analysis will be made in $\cal 
D$. In what follows, we assume that we are at a velocity $\O$ 
such that there is a vortex line and we want to find conditions 
on $\O$ and the elongation $\beta$ for the line to be stable and 
either straight or bent. 
  
First of all, it has been observed numerically \cite{GP1} that 
the vortex line lies in the plane closest to the axis of 
rotation and we can provide a rigorous justification: 
 
\begin{theo}\label{plane}  
If $\a\leq 1$, then the energy is minimized when the vortex line 
lies in the $(y,z)$ plane, that is the plane closest to the axis.  
\end{theo}  
Indeed, if we have a curve $\gamma$ parametrized as $\g(t)=(x(t),y(t),z(t))$, 
then we can define the new curve $\tilde{\g}(t)=(0,\ty(t),\tz(t))$ 
by $\tz(t)=z(t)$ and $\ty(t)=\sqrt{\a^2x^2+y^2}$. 
Then $\rtf({\g(t)})=\rtf({\tilde{\g}(t))}$. 
Since $\a<1$, $\dot{\ty}^2\leq\dot{x}^2+\dot{y}^2$, hence 
$\rho(\tg)|\dot{\tg}| - \Omega 
\rho(\tg)\dot{\tilde z} \le 
\rho(\g)|\dot{\g}| - \Omega \rho( \g)\dot{z}$. 
It follows that the 
energy of the new curve $E[\tg]$ is less or equal than ${E}[\g]$. If $\a=1$, that is the cross section is a disc, then our arguments imply that the vortex line is planar, but of course all transversal planes are equivalent.
 
From now on, we will assume  
that the curve lies in the plane $(y,z)$, 
so that $\rho$, given by (\ref{rho}), only depends on $y$ and $z$. 
Recall from the expression of $E$, (\ref{Eg}), that for $E[\g]$ to be negative, we need $\rho-\O\rho^2$ to be negative somewhere, that is $\Omega \rho > 1$. 
 For fixed $\O$, we define the regions   
\[   
\calD_g := \{ (y,z) : \Omega \rho(y,z) > 1 \},   
\quad   
\quad   
\calD_b := \calD \setminus \calD_g.   
\]   
We will   
refer to these sets as ``the good region'' $\calD_g$  
and  ``the bad region'' $\calD_b$ respectively. In the bad region,   
the energy of a vortex per unit arc length is necessarily   
positive, since $\rho-\O\rho^2>0$, whereas in the good region, for appropriately oriented   
vortices it can be negative since $\rho-\O\rho^2<0$.  One can see easily that for $\g$ to have a negative energy, part of the vortex line has to lie in the good region, that is close to the center of the cloud. Note that for $\calD_g$ to be non empty, we need at least $\O\r0>1$.  
In the region $\calD_g$, we will see that the vortex is close to the axis for all $\b$. On the other hand, in the region $\calD_b$, the vortex goes to the boundary along the quickest path: if $\b$ is small, perpendicularly to the boundary, which gives rise to a bent vortex and if $\b>1$, the vortex stays along the axis of rotation.   
   
The organization of the paper is the following: first  we study the local stability of the straight vortex:  
 if $\O$ is large, then the straight vortex is a local minimizer. That is when $\O$ gets large, the vortices tend to be straight, while if $\b$ is small then the straight vortex loses local stability and the first vortex to be nucleated is bent.  Then, we study the minimization of $E[\g]$ in $\calD_g$ and $\calD_b$ according to the value of $\b$.  
 We finally derive that a minimizer of the energy has a minimal length.   

\section{Stability and instability  
of the straight vortex}   
In this section, we  
study the stability of the straight vortex.  
We parametrize the straight vortex as $\gamma_s(z) = (0, z)$  
for $-\zmax < z < \zmax$, with $\zmax = \sqrt{\rho_0}/\beta$. One can compute $E[\gamma_s]$ and derive that it is 0 for $\O\r0=5/4$. 
We have two aims: first show that for $\b$ small, when the straight vortex has 0 energy or small negative energy, that is for $\O\r0$ close to $5/4$, then it is unstable. Then, we want to prove on the contrary that if $\b$ is fixed and $\O$ is sufficiently big, the straight vortex is stable.

 We consider  perturbations of the straight vortex of the  
form $\g_\delta(z) = (\delta v(z), z + \delta^2 w(z)) + O( \delta^3)$ for  
$|z| < \zmax$. We require that   
$w$ be chosen so that $\rho(\gamma_\delta(\pm\zmax)) = 0$, thereby respecting
the condition that the vortex line terminate at the boundary
of the cloud.
  
Writing a Taylor series expansion for $E$, one finds that  
\[  
E[\gamma_\delta] = E[\gamma_s] +\frac {\delta^2}2 (v, E''[\g_s]v) + O(\delta^3)  
\]  
where  
\[  
(v, E''[\g_s]v) =   
\int_{-\zmax}^{\zmax}  
2( 2\Omega \rho - 1)v^2 +  \rho {v'}^2 dz.  
\] 
Here and in the rest of this section, $\rho = \rho(0,z) 
= \rho_0 - \beta^2 z^2$. 
To get this it is necessary to integrate by parts and  
use the fact that the straight vortex solves the Euler-Lagrange equations  
for $E$. In particular this eliminates all terms involving $w$.  
No boundary terms arise from integration by parts  
because $\rho(\gamma_\delta) = 0$  
at the endpoints.  
  
We say that the straight vortex is stable if $(v, E''[\g_s] v) > 0$  
for all $v$, and unstable if $(v, E''[\g_s]v) < 0$ for some $v$.

\begin{theo}  
The straight vortex is stable if  
\[  
\Omega\rho_0 > \frac 34 + \frac 1 {4 \beta^2}.  
\]  
The straight vortex is unstable if $\beta < 1/\sqrt{3}$ and   
\begin{equation}  
\Omega\rho_0 < \frac 16 + \frac 1 {6 \beta^2} . 
\label{suff.unstable}\end{equation}  
 
\label{th.stable}\end{theo}

Note that the 2 values are consistent in the sense that they both scale 
like $1/\b^2$ when $\b$ is small.  
For $\O$ large, one expects several vortices in the condensate, but the fact that a 
 straight vortex is stable gives an indication that for $\O$ large, each vortex should be nearly straight, which is consistent with the observations 
 \cite{AK}.  
\begin{remark} 
It is interesting to see what happens in Theorem \ref{th.stable} when $\Omega\r0=5/4$, that is when the straight vortex has zero energy. The first inequality yields 
 that if $\beta > 1/{\sqrt{2}}$, then 
the straight vortex  is stable for all $\O$ such that $\O\r0>5/4$, that is when 
 $E[\g_s] < 0$. If $\b>1$, we will see that  
 $\g_s$ is not just stable but in fact minimizes $E$. 
The second inequality implies that, if $\beta < \sqrt{2/13} \approx .39$  
then the straight vortex is unstable at the velocity  
$\Omega \rho_0 = 5/4$ at which $E[\g_s] = 0$. As a result,  
for these values of $\beta$,   
the first vortex to nucleate as $\Omega$ increases  
is a bent vortex. Note that it has been observed in \cite{SF2} that for $\b<1/2$, the ground state of the system corresponds to a curved line. 
 
 All this indicates that by varying the elongation of the condensate, one may hope to go from a situation where the first vortex is bent, to a situation where it is straight. \end{remark} 
  
To prove the instability of the straight vortex, we will find  
explicit perturbations $v$ for which $(v, E''[\g_s]v) < 0$.  
These also indicate the shape of good test functions.  
  
We define a perturbation $v$ (depending on a parameter  
$\theta$, which for now we regard as fixed) by  
\[  
v(z) =  
\left\{ \begin{array}{ll}0	&\mbox{ if } z \le \theta \zmax\\	  
	\left( \frac z{\zmax} - \theta\right)(1-\theta)^{-1}  
&\mbox{ if }z \ge \theta \zmax.\end{array}\right.  
\]  
Here $v$ is normalized so that $v(\zmax) = 1$.   
For this choice of $v$, a lengthy but straightforward calculation  
shows that  
\begin{eqnarray*}   
(v, E''[\g_s]v)	  
	&&=   
\frac {2 \Omega \rho_0^{3/2}}{30 \beta}  
[(1-\theta)^2(\theta+4)\\  
&&\quad\quad- \frac 5{\Omega\rho_0}  
(1-\theta) - \beta^2(1 + \frac\theta 2)] \\  
	&&=:  
\frac {2 \Omega \rho_0^{3/2}}{30 \beta}  
\Delta(\theta).   
\end{eqnarray*}   
It follows that the straight vortex is unstable if  
\begin{equation}  
(1-\theta)^2(\theta+4) < \frac 5{\Omega\rho_0}  
\left((1-\theta) - \beta^2(1 + \frac\theta 2)\right)  
\label{unstable.cond}\end{equation}  
for some $\theta \in [0,1)$.  
It is helpful to write $\theta$ as $\theta = 1 -\eta \beta^2$  
for some $\eta>0$ to be determined. Then (\ref{unstable.cond})  
can be written in terms of $\eta$, as  
\[  
\Omega \rho_0 < 5\left(\frac{ 1 + (\beta^2/2) - (3/2\eta)}  
{\eta\beta^2(5-\eta\beta^2)}\right).  
\]  
This is satisfied if  
\[  
\Omega \rho_0 < \frac{ 1 + (\beta^2/2) - (3/2\eta)}  
{\eta\beta^2} = \frac 1{2\eta} + \frac 1{\eta\beta^2}(1  
- \frac 3{2\eta}).  
\]  
The extremum is achieved for $\eta$ close to 3, so we can take $\eta =3$  
to find that (\ref{suff.unstable}) is a sufficient condition for instability. 
Because $\theta = 1 - \eta \beta^2 \ge 0$, this conclusion only 
holds if $\beta \le 1/\sqrt{3}$. For larger values of $\beta$,  
one can make different choices of $\theta$ to find thresholds for 
instability.

To derive the sufficient condition for stability,  
note that for every $z$,  
\[  
\frac {3\rho}{2\rho_0} - \frac {(z \rho)'}{2\rho_0}  
= 1.  
\]  
Multiplying $v^2$ by the expression on the left and integrating  
by parts, we obtain  
\[  
\int_{-\zmax}^{\zmax}  v^2 dz   
=\int_{-\zmax}^{\zmax} \rho\left[ \frac{ 3v^2}{2\rho_0}   +  
\frac z {\rho_0}  v v'\right] dz.  
\]  
Since $|z|/\rho_0 \le \zmax/\rho_0 = 1/\beta\sqrt{\rho_0}$  
for $|z|<\zmax$,  
\[  
\int_{-\zmax}^{\zmax}  v^2 dz   
\le   
\int_{-\zmax}^{\zmax} \rho\left[ \frac 3{2\rho_0} v^2  
+ \frac1 {\beta\sqrt \rho_0}  |v|\ |v'| \right] dz.  
\]  
Now we use the inequality  
$a b \le a^2/2 +  b^2/2$,   
to deduce   
\[  
\int_{-\zmax}^{\zmax}  v^2 dz   
\le   
\int_{-\zmax}^{\zmax} \rho \left[( \frac 3{2\rho_0}  
+ \frac{1}{2\rho_0 \beta^2}) v^2 + \frac 12(v')^2\right] dz.  
\]  
In particular, if   
\[  
\Omega \rho_0 > \frac 34  + \frac 1{4\beta^2}  
\]  
then this implies that $(v, E''[\gamma_s] v) > 0$ for all $v$.  
This completes the proof of Theorem \ref{th.stable}.  
  
 \hfill 
 
We would like to derive a more precise estimate of the critical velocity for which a bent vortex minimizes the energy $E[\g]$. We have seen that for $E[\g]$ to be negative, we need at least $\O\r0>1$ so that the good region $\cd_g$ is nonempty. Note that $\O\r0=1$ is exactly the 2D critical 
velocity at the plane $z=0$ for the existence of a vortex. But  a bent 
vortex cannot be a minimizer of $E[\g]$ exactly at $\O\r0=1$, since the good region $\cd_g$ has to have 
some critical size so that the vortex energy in the good region provides a sufficient contribution to compensate the positive part due to the length 
in the bad region. On the other hand, for $\O\r0=5/4$, the straight vortex has 0 energy. Thus, the critical velocity to obtain a bent 
vortex is $1<\O_c\r0<5/4$.  
We want to obtain a sharper estimate by 
using appropriate test functions.    
To find good test functions, note that  
\[   
\Delta'(\theta) = [ 3 \theta^2 + 4 \theta   
-(7 - \frac 5{2\Omega\rho_0}(2+\beta^2)]   
\]   
and so $\Delta$ has a local maximum at   
\[   
\theta_* = -\frac23 + \sqrt {\frac {25}9 - \frac 5   
{6\Omega\rho_0}(1-\beta^2)}   
\]   
which lies in the interval $(0,1)$ for the parameter range   
that we care about.   
  

Note that $\theta_*$ is an increasing function of $\Omega$,   
which is  consistent with numerical calculations showing   
that for larger values of $\Omega$, the minimizing path stays   
close to the $z$ axis over a longer interval.  
For $\theta=\theta_*$, we compute the energy of  
the path which is straight between $z=-\theta$ and $\theta$ and  
goes to the boundary along a straight line. The optimal end point on the boundary  
is at $z=\theta+\b$ for $\b$ small. For this special test function $\g$, we can compute $E[\g]$ to find that it is less than 
$${{\O\r0}\over 8}  ({{53}\over 4}-36\b+21\b^2-4\b^3)-{{25}\over 8}+3\b-\b^2+10-8\O\r0$$ 
Thus,  for $\b$ small, we find an upper bound for the critical velocity which yields a negative energy for such a test function:    
$$\Omega \r0={{(220+96\b)}\over {(203+76\b)}}.$$   
In the condition of the ENS experiment, this yields $\O\r0<1.08$,  
that is in the original variable (see (\ref{O})), $\tO/\omega_y<0.385$, which is  
very close to the value found numerically 0.38 \cite{CM}.   
   
As a conclusion, we have shown that there is a critical value  
of $\Omega$ called $\Omega_c$ with 
$\Omega_c\r0 \approx 1.08$, such that a bent vortex has negative  
energy and less energy than a straight vortex.

\section{Shape of the vortex according to $\b$}   

In this section we prove that when the condensate cloud has a pancake shape, then the straight vortex 
is always minimizing among vortices with negative energy.

Recall that ${\cal D}=\{ \rho>0\}$ and we write    
$\g(t) = (y(t), z(t))$ to denote a generic vortex line  represented by   
a continuous Lipschitz function from  $I=[0,1]$ into   
$\overline{\calD}$ such that $\g(0), \g(1) \in \partial \cal D$.  
   
For such a curve $\g$,   
let $I_{\g,g} := \{ t\in I : \g(t) \in \calD_g \}$ and $I_{\g,b} = I\setminus   
I_{\g,g}$. And let $\g_g$ be the restriction of $\g(\cdot)$ to $I_{\g,g}$,   
and similarly $\g_b$.   
   
The definition of $I_{\g,b}$ implies that $\rho(\g(t)) - \Omega \rho^2(\g(t))>0$   
for $t\in I_{\g,b}$, and as a consequence   
\[   
\rho(\g(t)) |\dot \g(t)| - \Omega \rho^2(\g(t)) \dot z   
\ge |\dot \g(t)| \left( \rho(\g(t)) - \Omega \rho^2(\g(t)) \right)   
\]   
which is positive in $I_{\g,b}$. Thus if $\g$ is such that $I_{\g,g}$ is empty, then   
clearly $E[\g] > 0$ and it is energetically favorable not to  
have a vortex. This is the case in particular for $\O\r0<1$ since then  
$\calD_g$ is empty. We  
may thus restrict our attention to the case $I_{\g,b}$ nonempty.  
\begin{prop}   
Let $M_g = \inf \{ E[\g_g]\}$, where $\g_g$ is the restriction 
of $\g(\cdot)$ to $I_{\g,g}$, then for all $\b$ and all $\O$, the infimum is  
attained by the straight vortex.   
\label{prop1}\end{prop}   
\begin{prop}   
Let $M_b = \inf \{ E[\g_b]\}$, , where $\g_b$ is the restriction 
of $\g(\cdot)$ to $I_{\g,b }$, then for all $\b\geq 1$, the infimum  
is attained by the straight vortex.   
\label{prop2}\end{prop}   
Note that in the bad region, Proposition 2 only holds for $\b>1$.  
If $\b<1$, the situation is 
somewhat more complicated: $\int_{\gamma_b} \rho dl$ 
is minimized by a path that joins $\calD_g$ to $\partial \calD$ 
along the $y$ axis, whereas $-\int_{\gamma_b} \rho^2 dz$ is 
minimized by the straight vortex running along the $z$-axis. 
The minimizer of the full energy reflects the competition between  
these two terms, and hence is bent.  
  
We always  have   
\[   
E[\g] = E[\g_g] + E[\g_b] \ge  M_g +M_b  
\]   
In particular, as a corollary of the above  
propositions we deduce  
\begin{theo}For $\b\geq1$   
\beq \label{leg}   
E[\g]  \ge \inf (0,E[\g_s]) \eeq   
where $\g_s$ is the straight vortex along the $z$ axis. If $E[\g_s]< 0$, the   
equality in (\ref{leg}) can happen only if $\g$ is the straight vortex.   
\label{straight.opt}\end{theo}   
To prove Proposition \ref{prop1},  
first note that   
\[   
\int_{\g_g} \rho dl - \Omega \rho^2 dz \ge   
\int_{\g_g} \rho |dz| - \Omega \rho^2 dz \ge   
\int_{\g_g} (\rho - \Omega \rho^2) dz.   
\]   
Since we have assumed that $\g$ does not self-intersect,  
we can identify $\g$   
with the (oriented) boundary of an open set   
$V\subset \calD$. Then $\g_g$ can be identified with   
$\calD_g \cap \partial V = \partial(\calD_g \cap V)\setminus   
(\partial \calD_g \cap \bar V)$. Since $\rho-\Omega\rho^2 = 0$  
precisely on $\partial \calD_g$, this implies that  
\[   
\int_{\g_g} (\rho - \Omega \rho^2) dz   
\ = \    
\int_{\partial(\calD_g \cap V)} (\rho - \Omega \rho^2) dz.  
\]   
And by Stokes' Theorem,    
\[   
\int_{\partial(\calD_g \cap V)} (\rho - \Omega \rho^2) dz    
\ = \    
\int_{\calD_g \cap V} (1 - 2 \Omega \rho)\rho_y dy dz .   
\]   
The definition of $\calD_g$ implies that $1-2\Omega \rho < 0$,   
and so this integral is clearly minimized if   
$\calD_g \cap V$ is just the subset of $\calD_g$ where   
$\rho_y>0$, so that   
\begin{equation}  
\int_{\partial(\calD_g \cap V)} (\rho - \Omega \rho^2) dz    
\ \ge \    
\int_{\{ (y,z) \in \calD_g : y < 0 \}} (1 - 2 \Omega \rho)\rho_y dy dz .   
\label{sharp1}\end{equation}  
Again using Stokes Theorem and the fact that $\rho- \Omega \rho^2$   
vanishes on $\partial \calD_g$, we find that this is equal to  
\[   
\int_{-z_*}^{z_*} \Bigl ( \rho(0,z)  - \Omega \rho^2(0,z) \Bigr ) dz,  
\]   
where $(0,\pm z_*)$   are the points where the $z$-axis   
intersects $\partial \calD_g$.   
Combining these inequalities, we find that  
\begin{equation}   
\int_{\gamma_g} \rho \ dl - \Omega \rho^2 \ dz \ge   
\int_{-z_*}^{z_*} \Bigl ( \rho(0,z)  - \Omega \rho^2(0,z)\Bigr ) \ dz.  
\label{sharp2}\end{equation}   
It is easy to see that  
equality holds in (\ref{sharp1}),  
and hence in (\ref{sharp2}), exactly   
when $\gamma$ is the straight vortex, and so  
we have proved Proposition \ref{prop1}.

To prove Proposition \ref{prop2}, fix $\g $ such that $I_{\g,g}$ is nonempty.  
The beginning and end of $\g$ must lie in the bad region,   
and $\g$ intersects the good region, so $I_{\g,b}$ must  
consist of at least two components. Let $(a_1, b_1)$ denote  
the first such component and $(a_2, b_2)$ denote the  
last, and write $\g_1$ and $\g_2$ to denote the corresponding  
portions of $\g_b$,  
so that  $\g_1$ is parametrized as  
$\g_1 = (y,z): (a_1,b_1)\to \calD_b$, with  
$\g_1(a_1) \in \partial \calD$  
and $\g_1(b_1) \in \partial \calD_g$.  
We need to show that $\g_1$ and $\g_2$ both have more energy than  
the corresponding parts of the straight vortex. We will consider only  
$\g_1$, as the argument for $\g_2$ is exactly the same.  
  
Define $\g_s = (0, \zeta)$ to be a parametrization of the part of  
the straight vortex joining $(0, -\zmax)$ to $(0,-z_*)$,  
where $\zmax = \sqrt{\rho_0}/\beta$:  
\[   
\tilde \zeta(t) = -\frac 1\beta( y(t)^2 + \beta^2 z(t)^2)^{1/2},   
\   
\zeta(t) = \max_{a\le s \le t} \tilde \zeta(s).   
\]   
Recall that we have $\g_1=(y(t),z(t))$.  
The definition is arranged so that $t\mapsto \zeta(t)$  
is nondecreasing and $|\dot \gamma_s|=\dot\zeta$.  
To prove the proposition, it thus suffices to show that  
\[  
\rho(\g_1)|\dot \g_1| - \Omega \rho^2(\g_1) \dot z   
\ge   
\rho(\g_s)|\dot \g_s | - \Omega \rho^2(\g_s) \dot \zeta .  
\]  
If $\zeta(t) > \tilde\zeta(t)$, this is clear, because  
then $\dot \zeta = 0$, so the right-hand side vanishes while  
the left-hand side is nonnegative, by the defining property  
of the bad region $\calD_b$.  
  
And if $\zeta(t) = \tilde \zeta(t)$, then  
$\rho(\g_1(t)) = \rho(\g_s(t))$, and so in this case  
$0 \le  1 - \Omega \rho(\g_1(t)) =  
1- \Omega\rho(\g_s(t)) \le 1$. So we only need to show  
that  
  
\begin{equation}   
|\dot \g| - c \dot z \ \ge \  |\dot \g_s| - c \dot \zeta   
\label{claim2.2}\end{equation}   
for any $c\in [0,1]$. We will apply it to $c=\Omega\rho(\g_s(t))$.  
   
To do this, first note that    
\[   
\dot \zeta	   
= \dot {\tilde \zeta}	   
	=   
\frac 1{\tilde\zeta} \left(  \frac{y \dot y}{\beta^2} + z \dot z\right)   
	=(\dot y, \dot z) \cdot   
\left( \frac 1{\tilde \zeta}( \frac y{\beta^2}, z)\right) .  
\]   
So   
\[   
|\dot \zeta| \ \le \   
|\dot \g| \left(\frac 1{\tilde \zeta^2}(\frac{y^2}   
{\beta^4} + z^2)\right)^{1/2} =    
|\dot \g| \left(\frac { \beta^{-4}{y^2} + z^2}   
{\beta^{-2} y^2 + z^2}\right)^{1/2} .  
\]   
Since $\beta > 1$, we conclude that  
$|\dot \zeta| \le |\dot \g_1|$.  
Also, it is clear that $|\dot z| \le |\dot \g_1|$.   
So if $0\le \alpha \le 1$, then   
\[   
|\dot \g_1| - c \dot z \ge |\dot \g_1|(1-c) \ge   
\dot \zeta(1-c) = |\dot \g_s| - c \dot \zeta,   
\]   
which proves (\ref{claim2.2}), and hence Proposition \ref{prop2}. 
  

\section{Minimal length}   
In the case $\b<1$, that is when the vortex line is bent, we will  
prove that the vortex has a minimum length. This is related to the  
fact that the vortex has to go to the center of the cloud and 
spend some time in the good region.

For an open set $U\subset \mathcal{D}$ with    
Lipschitz boundary, we endow $\partial U$ with an   
orientation in the standard way, so that Stokes' theorem holds.   
   
We will prove the following isoperimetric-type inequality:   
   
\begin{theo} For every $0 < \beta \le 1$  
\begin{equation}  
\left|\int_{\partial U} \rho^2 dz \right|  \ \le \ ( 2\sqrt{\rho_0} )^{1/2}  
\left( \int_{\partial U} \rho dl \right)^{3/2}  
\label{isoperimetric.ineq}\end{equation}  
for every connected open subset $U \subset \mathcal{D}$,  
\end{theo}

\begin{remark} The exponent $3/2$ is the best possible. An inequality similar to (\ref{isoperimetric.ineq})   
is valid for $\beta>1$, but   
the proof needs to be modified a bit.   
 For the straight radial vortex,  
\[  
\int_{\partial U} \rho^2 dz  =  
\frac {16}{15}\frac{(\rho_0)^{5/2}}{\beta}  
\quad  
\mbox{ and }  
\quad  
\int_{\partial U} \rho dl = \frac 43\frac{(\rho_0)^{3/2}}{\beta},  
\]  
and so  
\[  
\left(\int_{\partial U} \rho^2 dz\right)  
\left(\int_{\partial U} \rho dl\right)^{-3/2}  
\approx  
0.52\  \beta^{1/2}(\rho_0)^{1/4}.  
\]  
This shows that the constant $(2\sqrt{\rho_0})^{1/2}$ in   
(\ref{isoperimetric.ineq}) is fairly close to sharp  
for $\frac 14 \le \beta < 1$ say.  
\end{remark}

1. We use Stokes' Theorem to calculate  
\[  
\int_{\partial U} \rho^2 dz  
= 2 \int_U \rho \rho_y dy dz  
\le 2 \int_{U^-} \rho \rho_y dy dz  
\]  
where $U^- = \{ (y,z) \in U \ : \ y<0 \}$, since $\rho \rho_y \le 0 $  
for $(y,z)\in \mathcal{D}$ such that $y\ge 0$.  
  
So the coarea formula implies that  
\begin{eqnarray*}  
&&\int_{\partial U} \rho^2 dz  
        \le  
2 \int_{U^-} \rho\frac{|\rho_y|}{|\nabla\rho|}  |\nabla \rho| dy dz     \\  
        &&=  
2 \int_{\rho_*}^{\rho^*} s  
\left( \int_{\{(y,z) \in {U^-} \ :  \ \rho(y,z) = s \} }  
\frac{|\rho_y|}{|\nabla\rho|}  
\ dl \right)ds  
\end{eqnarray*}  
where $ \rho_* = \inf \{ \rho(y,z) \ : \ (y,z)\in U \}$,   
and $\rho^* = \sup \{ \rho(y,z) \ : \ (y,z)\in U \}$.  
Thus  
\begin{eqnarray*}  
&&\left| \int_{\partial U} \rho^2 dz \right|  
\ \le \  
|\rho^* - \rho_*|\\  
&&\quad  
\sup_s \left( s  
\int_{\{(y,z) \in U \ :  \ \rho(y,z) = s \} }  
\frac{\rho_y}{|\nabla\rho|}\  dl 
\right).  
\end{eqnarray*}  
Thus to prove the theorem it suffices to establish the following  
two claims:  
\begin{equation}  
s\int_{\{(y,z) \in U \ :  \ \rho(y,z) = s \} }  
\frac{\rho_y}{|\nabla\rho|} dl  
\ \le \   \int_{\partial U}\rho dl  
\label{claim1}\end{equation}  
for every $s$, and  
\begin{equation}  
|\rho^* - \rho_*| \le (2\sqrt{\rho_0})^{1/2}  
\left( \int_{\partial U} \rho dl \right)^{1/2}.  
\label{claim2}\end{equation}

2. We first prove (\ref{claim1}). Fix some $s\in (\rho_*, \rho^*)$  
and write $\Gamma_s$  to denote  
$\{(y,z) \in {U^-} \ :  \ \rho(y,z) = s \}$. Also, let $\tilde \Gamma_s$  
denote $\partial U \cap \{ \rho \ge s\}$.  
  
First assume for simplicity that $\Gamma_s$ is connected, so that  
it consists of the short arc of the ellipse $\{ \rho = s \}$  
joining two points, say $p_0 = (y_0, z_0)$ and $p_1 = (y_1, z_1)$  
with $z_0 < z_1$.  
We can represent $\Gamma_s$ as the image of the mapping  
\[  
z\mapsto (y(z), z) = ( -(s - \beta^2 z^2)^{1/2}, z),  
\quad\quad  
z_0 < z < z_1.  
\]  
Differentiating the identity $\rho(y(z), z) = s$ we find that  
$\rho_y y'(z) + \rho_z = 0$. Thus  
\[  
\left| \frac d{dz} ( y(z), z)\right|  
=  
\left( 1 + y'(z)^2\right)^{1/2}  =  
\left( \frac{(\rho_y^2 + \rho_z^2)}{\rho_y^2}\right)^{1/2}  
 = \frac{|\nabla\rho|}{|\rho_y|}.  
\]  
It follows that  
\[  
s\int_{\{(y,z) \in U \ :  \ \rho(y,z) = s \} }  
\frac{\rho_y}{|\nabla\rho|} dl 
\ = \  
s\int_{z_0}^{z_1} dz.  
\]  
On the other hand, the one-dimensional measure of  
$\tilde \Gamma_s$ is certainly greater than $|p_1 - p_0| \ge  
z_1 - z_0$, and $\rho \ge s$ on $\tilde \Gamma_s$, and  
so  
\[  
\int_{ \tilde\Gamma_s } \rho(z,y) dl  
\ \ge \  
s\ l( \tilde \Gamma_s) 
\ \ge \  
s(z_2 - z_1).  
\]  
This proves (\ref{claim1}) if $\Gamma_s$ is connected. If not, one can  
apply the same argument on each connected component of  
$\Gamma_s$.  
  
3. Next we prove (\ref{claim2}). Let $q_*$ and $q^*$ be points  
in $\partial U$ such that $\rho(q_*) = \rho_*$, $\rho(q^*) = \rho^*$.  
Since we have assumed that $U$ is connected,  $\partial U$ contains  
a path joining $q_*$ to $q^*$. In fact it contains two such paths.  
If we write $\mathcal{P}$ to denote the set of all  
Lipschitz paths in $\mathcal{D}$ joining the level set  
$\{ \rho = \rho_*\}$ and the level set $\{\rho = \rho^*\}$,  
it follows that  
\[  
\int_{\partial U}\rho \ dl \ \ge 2 \ \inf_{\gamma\in \mathcal{P}}  
\int_\gamma \rho \ dl .  
\]  
Arguments in the proof of Proposition \ref{prop2} show that  
for  $\beta \le 1$,  
$\inf_{\gamma \in\mathcal{P}}\int_{\gamma}\rho dl$  
is attained by a path that goes in a straight  
line along the $y$ axis.  
Thus  
\[  
\inf_{\gamma\in \mathcal{P}} \int_\gamma \rho \ dl =  
\int^{y_*}_{y^*} (\rho_0 - y^2) \ dy,  
\]  
where   
$y_* = \sqrt{\rho_0 - \rho_*},  
y^* = \sqrt{\rho_0 - \rho^*}$.  
And since $y_*, y^* \le \sqrt \rho_0$,  
\begin{eqnarray*}  
\int^{y_*}_{y^*} (\rho_0 - y^2) \ dy  
&& \ge  
\frac 1{2 \sqrt{\rho_0}}  
\int^{y_*}_{y^*} (\rho_0 - y^2) 2y \ dy \\  
&& = \  
\frac 1{2 \sqrt{\rho_0}}  
\int_{\rho_*}^{\rho^*} \rho d\rho \\  
        &&= \  
\frac 1{4 \sqrt{\rho_0}} \left((\rho^*)^2 - (\rho_*)^2\right).  
\end{eqnarray*}  
Since $b^2 - a^2 \ge (b-a)^2$ when $0<a<b$, we deduce that  
(\ref{claim2}) holds. This concludes the proof of the theorem.  
  
 A short calculation starting from (\ref{isoperimetric.ineq}) 
shows that if $E[\g]<0$ then 
\beq\label{lb} 
\int_\g \rho dl >  {1\over {(2 \Omega^2 \sqrt{\rho_0})}}. 
\eeq 
We expect that even for a configuration with multiple vortices, adding a new vortex would require to overcome an energy barrier which implies that the length of the new vortex has a lower bound of the type (\ref{lb}).

   
\section{Conclusion}   
We have studied the shape of the first vortex line to be nucleated in a harmonic anisotropic rotating potential, 
according to $\O$ and the elongation of the cloud $\b$. We investigate the stability of the straight vortex and get that when $\O$ is large, the straight vortex is a local minimum of the energy. 
 We prove that when a vortex is nucleated,  
it is close to the axis
of rotation where the condensate density is high, and that near the
boundary, where the density is low, the shape of the vortex depends on
whether the cloud has a cigar or pancake shape.  
This shape reflects the competition in the energy between the rotation and the inhomogeneity of the trap, which makes the geometry of the experiment very important. 
In the case $\b>1$ (pancake), the vortex stays straight along the $z$ axis while in the case $\b$ small (cigar), the vortex is bending. In the case $\b$ small, this allows us to define an energy barrier for the nucleation of vortices and to prove that when a vortex line is nucleated, it has to have a minimal length.

\end{document}